\documentclass[reprint,5p,10pt,times,sort&compress,twocolumn]{elsarticle}
\usepackage{amsfonts}
\usepackage{amsmath} % math
\usepackage{amssymb} % math
\usepackage{graphicx}
\usepackage{color}
\usepackage{subfigure}
\usepackage{lineno}
\journal{Optik}
\begin{document}
\begin{frontmatter}
 
\title{Chaos-based image encryption using vertical-cavity surface-emitting lasers}
\author[AR]{Animesh Roy}
\ead{ aroyiitd@gmail.com}
\author[AR]{A. P. Misra\corref{cor1}}
\ead{apmisra@visva-bharati.ac.in, apmisra@gmail.com}
\address[AR]{Department of Mathematics, Siksha Bhavana, Visva-Bharati University, Santiniketan-731 235, West Bengal, India}
%\address[AR]{Department of Mathematics, Siksha Bhavana, Visva-Bharati University, Santiniketan-731 235, West Bengal, India}
\author[SB1,SB2]{Santo Banerjee}
\address[SB1]{Malaysia-Italy Centre of Excellence for Mathematical Sciences, Universiti Putra Malaysia, Malaysia}
\address[SB2]{Institute for Mathematical Research, Universiti Putra Malaysia, Selangor, Malaysia }
\ead{santoban@gmail.com}
\cortext[cor1]{Corresponding author.}

\begin{abstract}
 We study the encryption and decryption processes of color images using  the  synchronization of polarization  dynamics in a free-running vertical-cavity surface-emitting laser (VCSEL). Here, we consider a  bidirectional  master-slave configuration or two-way coupling with two  VCSELs. The latter are shown to exhibit hyperchaos and  synchronization with a high level of similarity between their emission characteristics.    The coupled VCSELs are then used as a transmitter and  a receiver for the communication of image or data. Furthermore, we propose a modified chaos-based image encryption algorithm using the pixel- and bit-level permutations which provides robust, faster and simpler encryption/decryption compared to many other chaos-based cryptosystems. The performances of the new cryptosystem   are analyzed and compared with a recently developed scheme [Opt.  Las.  Eng. 90 (2017) 238-246].      The security analysis and  some statistical investigations   show  that  the proposed cryptosystem is resistant to various types of attacks and  is efficient for secure communications in nonlinear optical media.
\end{abstract}

\begin{keyword}
Image encryption \sep Bit-level permutation \sep Surface-emitting laser \sep Chaos \sep Synchronization 



\end{keyword}

\end{frontmatter}

\section{Introduction} \label{sec-introduct}
The advancement of public  communication systems, such as satellite, mobile-phone, computer networking, Internet etc., has  led to vulnerability in secure communication of e.g., the  transmission of  confidential data like military data, confidential videos, messages  etc. In this way, the theory of cryptography has been developed (For some recent works, see, e.g., Refs.\citep{xu2017,li2017,assad2016,xiao2016,xu2014,virte2013,banerjee2011,assad2016,daneshgar2015,nosrati2017}).  On the other hand, the invention of semiconductor laser diodes, e.g., the vertical-cavity surface-emitting lasers (VCSELs) has been gaining its potential applications in laser devices considering their numerous advantages over  Light Emitting Diode (LED) and Edge Emitting Laser (EEL),  such as low threshold, circular beam profile, and on-wafer testing capability \cite{iga2013,panajotov2012}. Given their electro-optical characteristics and ability to modulate at frequencies ($\gtrsim25$ Gbps), VCSELs are ideal for high-speed communications and precision sensing applications.  They are also used for reliable operation at distances ranging from very close proximity links (i.e., centimeters) up to $500$ m in data center, enterprise, and campus networks. Furthermore, 
such VCSELs have been widely used in data communication industry for more than $15$ years serving in  data infrastructure links including $10$, $40$ and $100$ Gbps Ethernet, $16$ Gbps fiber channel, and $10$, $14$, and soon $25$ Gbps lane Infinite Band.  VCSELs are also emerging as an enabling technology across a wide range of applications, including touchless sensing, chip-to-chip interconnect, and gesture recognition.  

 It has been shown that   VCSELs can also exhibit nonlinear polarization dynamics and chaos \cite{virte2013}. Such chaos can be obtained in a number of ways, e.g., when the lasers are driven (into chaos) due to optical feedback from an external reflector. Furthermore, optical lasers like   VCSELs or semiconductor lasers are   used as secured media for transmission of confidential data, videos, messages etc.  These lasers are also  used for encryption-decryption of color images in the context of  chaos based cryptography \cite{banerjee2011}. 
\par
It is to be noted that two identical but independent chaotic systems cannot exhibit the same behaviors unless it is coupled or linked in some ways. In the latter, the system's evolution becomes identical, which is known as the chaos synchronization. Such exciting property of a dynamical system led to the development of secure chaos communication systems where the sender hides a message within the chaotic signal that can   only be recovered by the receiver at the synchronized state.  This approach has been applied in many secure communications, especially in optical chaos communication systems because of the added security and the speed of optical communications \cite{banerjee2011,thang2016}.   
 \par  In classical cryptographic schemes (e.g.,  AES, DES, One time pad), public key cryptography is  widely used for secure networking system. However,  these schemes  have some limitations in fast encryption on large data scales, such as those in color images, videos or audio data etc.  These are not only sequences of large data sets, but also each sequence is highly correlated with another. Encryption of these data set with the classical schemes, as above,  takes a longer time and thereby makes the system much slower (see, e.g., Ref. \cite{subramanyan2011}). In order to resolve this issue, many authors have proposed   chaos based cryptography schemes \cite{xu2017,li2017,assad2016,socek2005,xiao2016} in which a  nonlinear dynamical system, which exhibits chaos, is considered for encryption and decryption \cite{virte2013, banerjee2011}.  
 On the other hand, the data encryption in chaotic medium is known to be much efficient   than the traditional method in which   it is more easier for   hackers to recover the confidential data.  Here,  we consider a RGB color image which is a large set of data  and its color distribution is highly correlated with the data set. So, although the transmission of  these kind of data using the traditional encryption scheme  is secured but   security is much enhanced if we use a non-pattern medium like  chaotic medium.
  \par In this work,  we consider a quantum spin-flip  model (SFM) of VCSELs \cite{virte2013,hermier2002,martin1997,michalzik2003}, to be given in Sec. \ref{sec-model},  which  is used for encryption and decryption of a RGB image using a modified  chaos based cryptography scheme.   It is shown that the coupled VCSELs can exhibit hyperchaos and synchronization with a wide range of values of the parameters. A new hyperchaos-based image encryption algorithm using the pixel- and bit-level permutations, which modifies the previous one \cite{li2017}, is proposed and tested with an RGB image.     It is seen that the new cryptosystem is robust, faster, simpler  and more secured in comparison with Ref. \citep{li2017} and other chaos-based cryptosystems \cite{xu2017,li2017,assad2016,socek2005,xiao2016,xu2014}.  A statistical  investigation   is also carried out to ensure that the proposed encryption scheme   is free from any brute force attack.  
\section{The  model of VCSELs and their chaotic properties} \label{sec-model}
%\subsection{The model} 
We consider the nonlinear dynamics of   right- and left-circularly polarized (RCP, LCP) emission arising from  the recombination of two distinct carrier populations $D_+$ and $D_-$ in VCSELs. The latter have  
 a high quantum efficiency and low threshold which can operate on a very high rate optical communication in the range of several GHz. In terms of the slowly varying electromagnetic (EM) fields $E_{\pm}$ (normalized by the equilibrium value $E_0$) for RCP and LCP emission, we have the following set of equations \cite{virte2013,martin1997,michalzik2003}
\begin{equation}
\frac{dE_{\pm}}{dt}=\kappa(1+i\alpha)(N \pm n -1)E_\pm-i\gamma_pE_\mp - \gamma_aE_\mp, \label{E-m}
\end{equation}
\begin{equation}
\frac{dN}{dt}=-\gamma(N-\mu)-\gamma\left[(N+n)|E_+|^2+(N-n)|E_-|^2\right]|E_0|^2, \label{N-m}
\end{equation}
\begin{equation}
\frac{dn}{dt}=-\gamma_s n-\gamma\left[(N+n)|E_+|^2-(N-n)|E_-|^2\right]|E_0|^2, \label{n-m}
\end{equation}
where $N,n=D_+\pm D_-$ are the normalized carrier populations, $\kappa$ is the decay rate of the electric field in the cavity, $\alpha$ is the linewidth enhancement factor, and $\mu$ is the normalized injection current. Furthermore, $\gamma$ is the carrier decay rate, $\gamma_s$ is the spin-flip relaxation rate which models the process allowing the equilibration of the carrier population between the two reservoirs, and $\gamma_p$ and $\gamma_a$ are, respectively, the phase and amplitude anisotropies inside the laser cavity.    
\par
\begin{figure*}[hbtp]
 \centering
 \includegraphics[scale=.4]{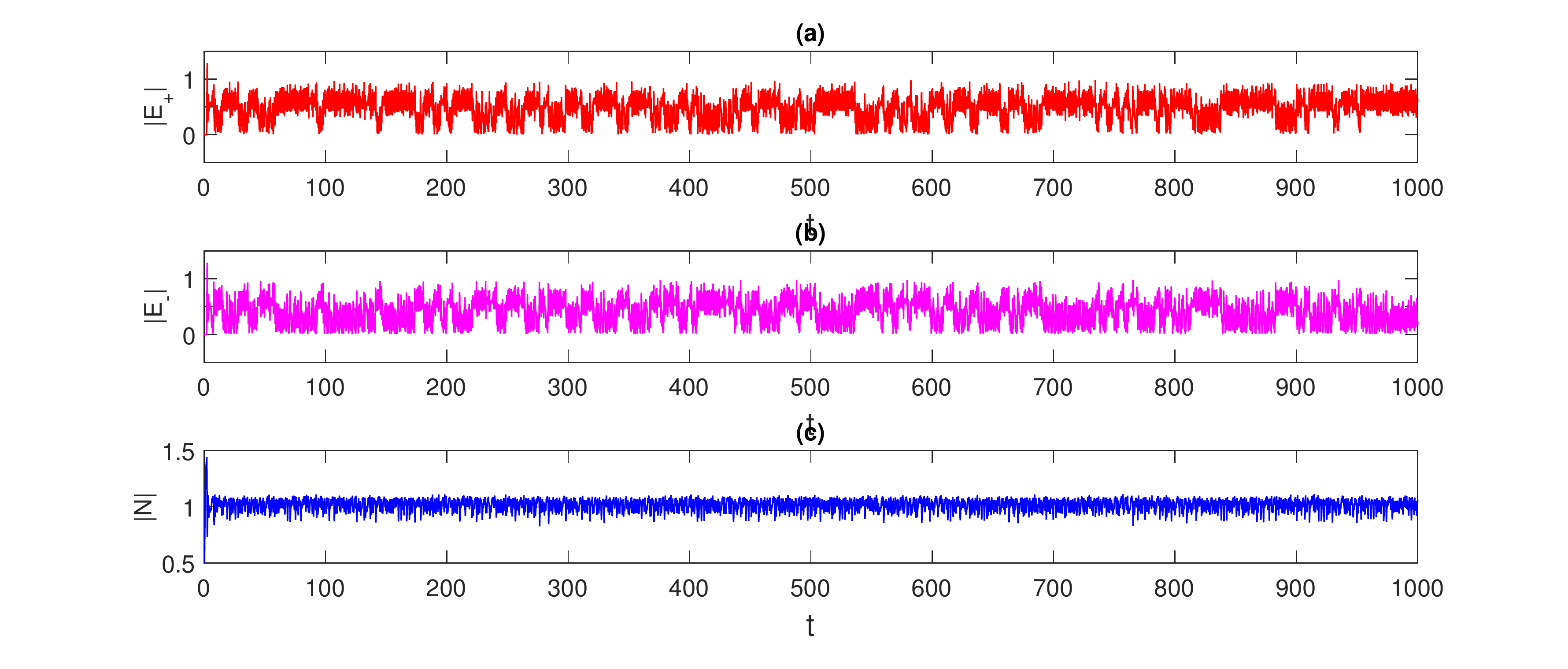}
 \caption{Numerical solution of Eqs. \eqref{E-m}-\eqref{n-m} for the forward and backward electric fields ($E_{\pm}$,   the upper and middle  panels) and the carrier population density ($N$, the lower panel) which exhibit chaos.}
 \label{fig:chaos-master}
 \end{figure*}

\begin{figure*}[hbtp]
 \centering
 \includegraphics[scale=.4]{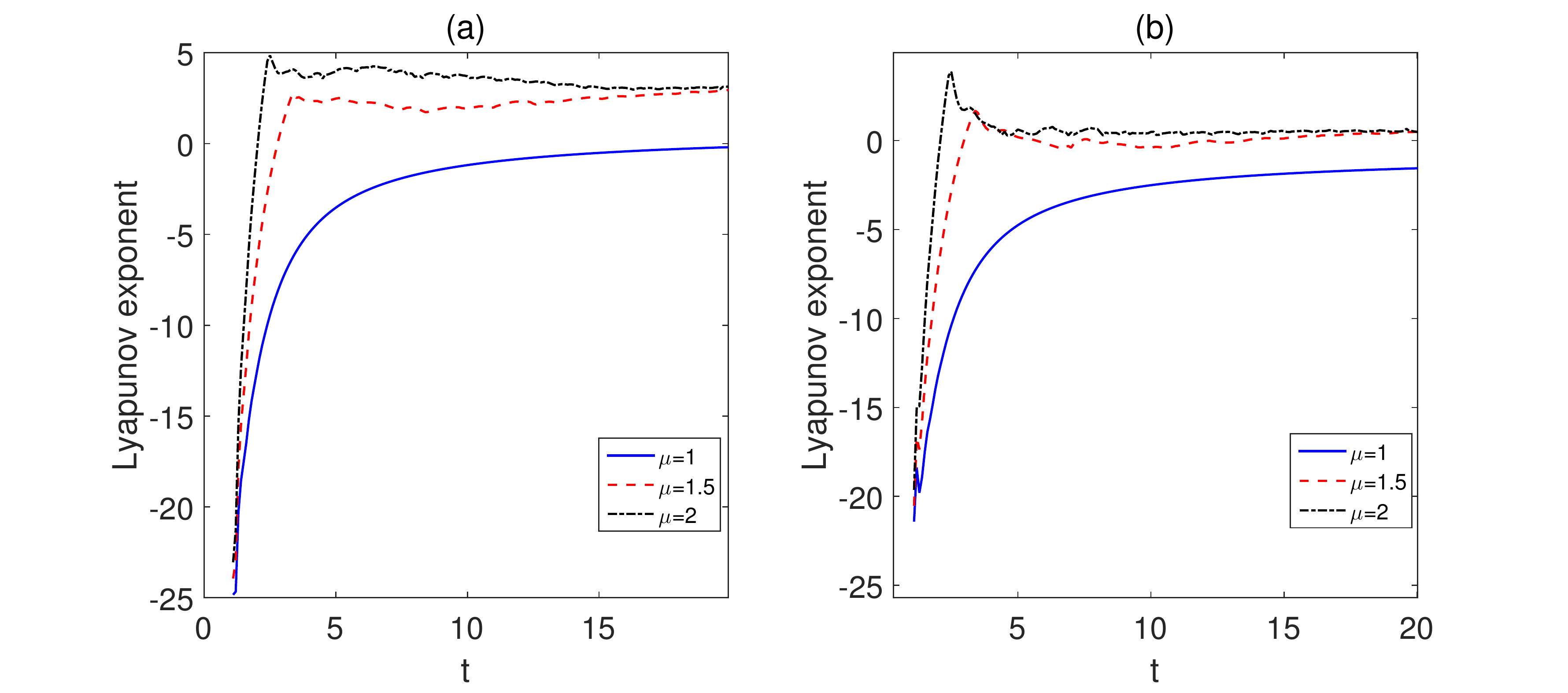}
 \caption{The two Lyapunov exponent spectra, corresponding to the system of Eqs. \eqref{E-m}-\eqref{n-m}, are shown with time $t$ in the subplots (a) and (b) for different values of the injection current parameter $\mu$. It is seen that as $\mu$ increases, the Lyapunov exponents turn over from negative to positive values.   The other two exponents   are always negative (not shown) irrespective of the values of $\mu$.   The  parameter values are taken as $\kappa=26$ ns$^{-1}$, $\alpha=3,~\gamma_s=5,~\mu=1.5$, $\gamma_p=5$ ns$^{-1}$ and $\gamma_a=-7$ ns$^{-1}$.  }
 \label{fig:lyapunov}
 \end{figure*}

 
In order to establish chaos, we numerically solve the system  of Eqs. \eqref{E-m}-\eqref{n-m} by a fourth order Runge-Kutta scheme with a time step size $t=0.01$ and an initial conditions $E_{\pm}=0.001,~ N=0.003,~ n=0.001$. The typical parameter values are considered as 
\begin{itemize}
\item $1\leq\kappa\leq100$  n$s^{-1}$, $2\leq\kappa_{inj}\leq10$  n$s^{-1}$,  $\alpha=3$, $2\leq\Delta\leq10$  n$s^{-1}$, 
\item $0\leq\gamma_p\leq100$  n$s^{-1}$, $-7\leq\gamma_a\leq7$  n$s^{-1}$, $1.45\leq\gamma\leq1.5$  n$s^{-1}$, $0\leq\gamma_s\leq100$  n$s^{-1}$.
\end{itemize}
The results are displayed in Fig. \ref{fig:chaos-master} after the end of the simulation at $t=1000$. From Fig. \ref{fig:chaos-master}, it is seen that both the polarized electric fields $E_{\pm}$ of the master laser exhibit chaos along with the carrier population densities $N$ and $n$.  It is found that the chaotic state of the system can be reached due to   the increasing values of the injection current parameter $\mu$.  
%%%%%%%%%%%%%%%%%%%%%%%%%%%%%%%%% Chaos   %%%%%%%%%%%%%%%%%%%%%%%%%%%%%%%%%
 Furthermore, in order to have some confirmation of our results, we have computed the   largest Lyapunov exponents as exhibited in  Fig. \ref{fig:lyapunov} with the same parameter values  as for Fig. \ref{fig:chaos-master}.   It is found that of the four exponents, two are always negative (not shown in the figure), and two others may be positive or negative depending on the values of the injection current parameter $\mu$.  From Fig. \ref{fig:lyapunov}, it is evident that the two lyapunov exponents can turn over from negative to positive values as the values of $\mu$ increase,  leading to chaos (more specifically hyperchaos)  for a longer time. This is in consequence  with the fact that the chaos in VCSELs is obtained  when the lasers are  driven due to  the optical feedback from an external reflector. 
 \section{Synchronization of master and slave VCSELs}\label{sec-chaos-synchro}
 We investigate the synchronization of two nearly coupled VCSELs, namely the master and the slave lasers which exhibit chaos as in Sec. \ref{sec-model}. We call the model  Eqs. \eqref{E-m}-\eqref{n-m} as the master VCSEL and couple it with   another VCSEL model, called the slave VCSEL, which   is very similar to the master one except with some coupling terms $\propto \kappa_{inj}$. Thus, the equations for the slave VCSEL model are
\begin{eqnarray}
\frac{dE_{s\pm}}{dt}&=\kappa(1+i\alpha)(N_s \pm n_s -1)E_{s\pm}-i\gamma_pE_{s\mp} \nonumber\\
&- \gamma_aE_{s\mp}- \Delta E_{s\pm}+\kappa_{inj} E_{\pm},\label{E-s}
\end{eqnarray}
\begin{eqnarray}
\frac{dN_s}{dt}&=-\gamma(N_s-\mu)-\gamma\left[(N_s+n_s)|E_{s+}|^2 \right.\nonumber\\
&\left.+(N_s-n_s)|E_{s-}|^2\right]|E_0|^2, \label{N-s}
\end{eqnarray}
\begin{eqnarray}
\frac{dn_s}{dt}&=-\gamma_s n_s-\gamma\left[(N_s+n_s)|E_{s+}|^2\right.\nonumber\\
&\left.-(N_s-n_s)|E_{s-}|^2\right]|E_0|^2, \label{n-s}
\end{eqnarray}
where $E_{s\pm}$ are the electric fields,   $N_s$ and $n_s$ are the total and difference of carrier population densities of the slave laser,    $\Delta$ is the detuning pulsation (difference between the frequencies of the master and slave lasers at the threshold of no anisotropy)  and $\kappa_{inj}$ is the coupling coefficient. Similar to Fig. \ref{fig:chaos-master}, the system of Eqs.   \eqref{E-s}-\eqref{n-s} for the slave laser can also be shown to exhibit hyperchaos with the same set of values of the parameters. 
\par 
The essential prerequisite for the   synchronization of the two systems is that the  VCSELs  operate in a chaotic regime when subject to an optical feedback.  In the case of very low optical coupling, i.e., with a lower values of the coupling coefficient $\kappa_{inj}$, the correlation between the outputs of the master and slave lasers is rather poor. However, as the coupling is enhanced with an increased value of  $\kappa_{inj}$  to its optimum value, the correlation is significantly improved and   the lasers are said to be synchronized.  Further increase in the coupling coefficient may result into ``no synchronization".  In the former case, it is apparent that the signal from the master laser is too weak to affect the synchronization, whereas in the latter,  the behavior of the slave is too strongly affected by the master laser at the largest value of the coupling coefficient. 
\par
\begin{figure*}[hbtp]
 \centering
 \includegraphics[scale=.5]{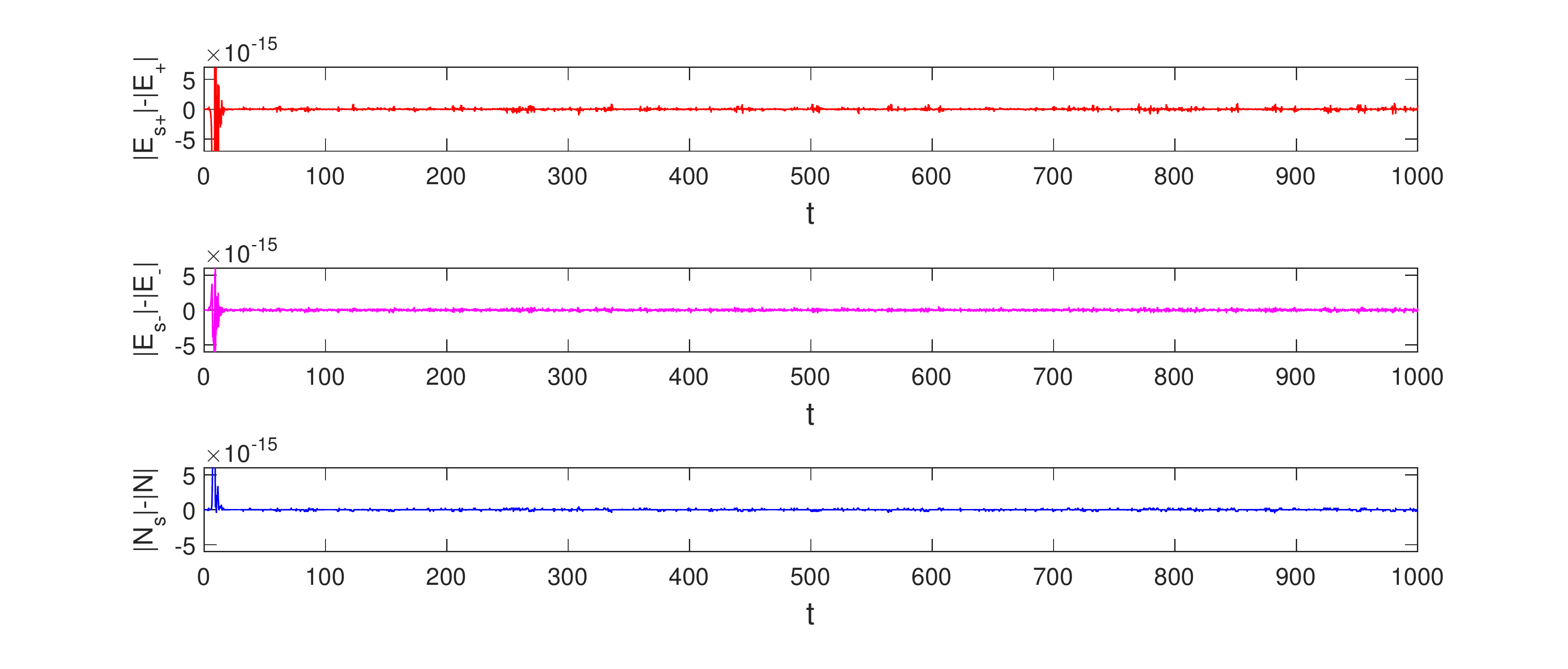}
 \caption{Synchronization errors of the electric fields and the density populations of the coupled systems of VCSELs given by Eqs. \eqref{E-m}-\eqref{n-m} and Eqs. \eqref{E-s}-\eqref{n-s}}
 \label{fig:synchro}
 \end{figure*}

Next, we study the synchronization properties of the   master and   the slave  lasers (with suffix $s$) given by Eqs.  \eqref{E-m}-\eqref{n-m} and \eqref{E-s}-\eqref{n-s} which exhibit chaos. These two systems, in fact, describe the driver and response systems for communications in chaotic media through synchronization in VCSELs.  In Eq. \eqref{E-s} we have introduced a coupling term with a coefficient $\kappa_{inj}$ which behaves as a noise-like term.   
A numerical simulation with an initial condition as for Fig. \ref{fig:chaos-master} reveals that synchronization of the two systems is, indeed, achieved after a certain time through the coupling term $\propto \kappa_{inj}$ (See Fig. \ref{fig:synchro}). The   parameter values taken are $\Delta=\kappa_{inj}=10$ n$s^{-1}$. The other parameter values are as for Fig. \ref{fig:chaos-master}.  Basically, Fig. \ref{fig:synchro}   displays the synchronization error between the driving and  response lasers at the most chaotic state. It is seen that the corresponding errors for the electric fields and the carrier densities are of the orders of $10^{-15}$. Thus, at the state of synchronization the system of Eqs. \eqref{E-m}-\eqref{n-m} can be considered as transmitter and the system of Eqs. \eqref{E-s}-\eqref{n-s} as the receiver one.
%%%%%%%%%%%%%%%%%%%%%%%%%% Synchronization  %%%%%%%%%%%%%%%%%%%%%%%%%%
 %%%%%%%%%%%%%%%%%%%%%%%%%%%%%%%%%%%%%%%%%%%%%%
%%%%%%%%%%%%%%%%%%%
\section{Transmission of data through chaotic medium}\label{sec-data-transm}
\begin{figure*}[hbtp]
 \centering
 \includegraphics[scale=1.2]{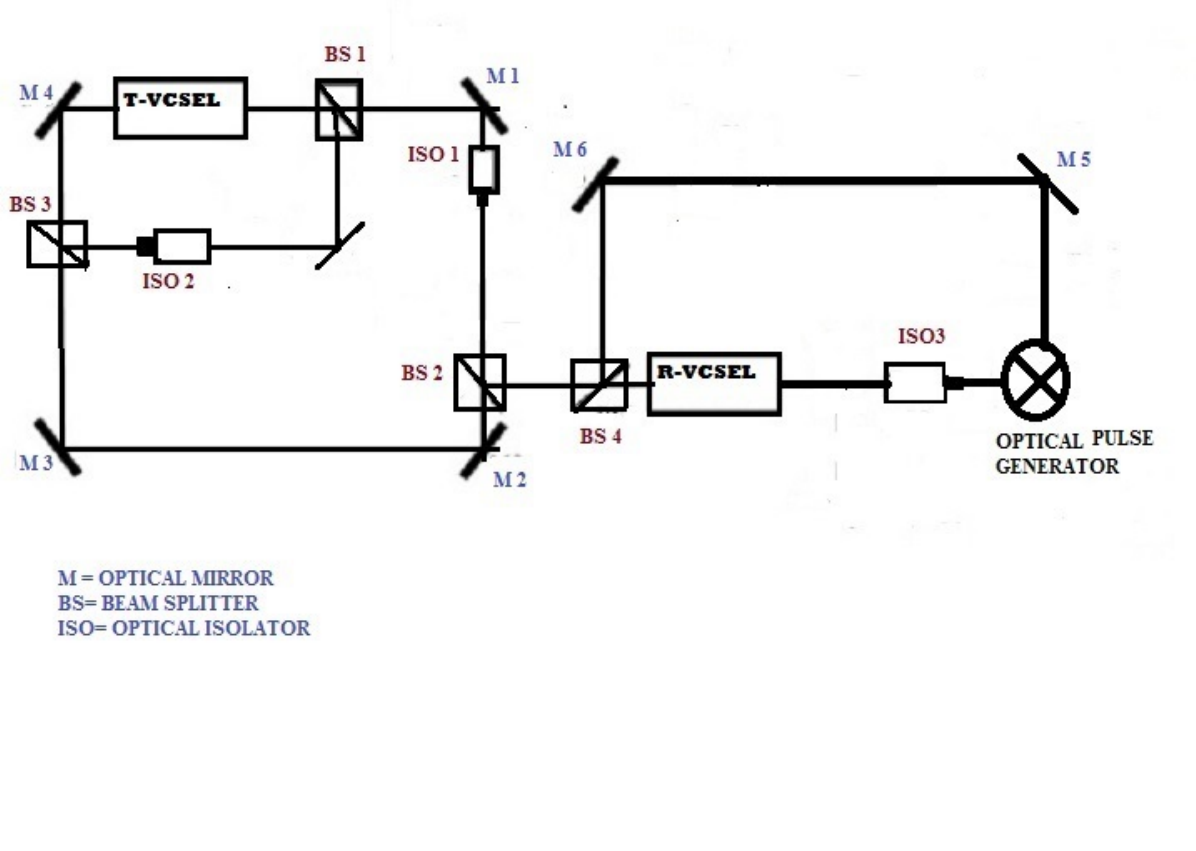}
 \caption{A schematic diagram for the coupling of VCSELs used for data transmission}
 \label{fig:schema}
 \end{figure*}

We consider the coupling between two systems of VCSELs   for transmission of  data. A schematic diagram is given in Fig. \ref{fig:schema} to demonstrate how a high resolution  data can be transferred from the transmitter (T-VCSEL) to the receiver (R-VCSEL) by means of synchronization of these coupled systems.  However,  the main concern is that each laser has its own noise which may prevent establishing the synchronization. In order to get rid of this situation, i.e., to reduce the noise from the systems, the optical beam splitter and the optical isolator are used. It has been shown   that  the VCSEL model  has   lesser noise emission   than the semiconductor lasers \cite{michalzik2003}.  The coupling of the two VCSELs and the  transmission of  data   through chaotic medium are demonstrated step by step as follows:
 \begin{itemize}
 \item  Take an RGB image (data) and switch on the transmitter system. The  current   flows through the laser diode.
\item  Chaos sequence is formed   of the  electric fields $E_{\pm}$ and the carrier densities $N,~n$.
\item  The master laser reads the image pixels as a sequence of data. 
\item  Relate the two sequences of data set, namely the chaotic data   and the   pixel data   with, e.g., an XOR operation.  This part is called the encryption.
\item   Switch on the receiver system  (slave laser) and synchronize it with   the master laser.  Data-pulse then  propagates from the master   to  the slave laser and     chaos is established in the slave laser.
\item   Arrange the chaotic data sequences of the slave laser, and use it to decrypt the image.
\end{itemize}
 %%%%%%%%%%%%%%%%%%%%%%% Diagram-transmitter- Receiver   %%%%%%%%%%%%%%%%%%%%%%%%%%%
 %%%%%%%%%%%%%%%%%%%%%%%%%%%%%%%%%%%%%%%%
 In Sec. \ref{sec-encryption}, we will discuss these steps in more detail. 
\section{Chaos-based image encryption and decryption} \label{sec-encryption}
In this section we propose an algorithm for encryption of a high quality RGB image. In the latter,    we have a three-dimensional matrix  $P[m,l,k]$ and all the elements are   integer  modulo   $256$,  i.e., they lie in  between $0-255$ and are  called the voxel values of the image. The color distribution is  formed of red (R), green (G) and blue (B). We follow the similar process as in Ref. \citep{li2017,banerjee2011}  for the encryption of an RGB image. First, we generate the key vectors by the chaotic system (master laser) and reshape the corresponding matrix as per the size of the image. Then a pixel-level permutation is employed to shuffle the pixel positions of the image. Next,  we change the bits of each pixel value  of the shuffled sequence by the bit-level permutation matrix  in order  to strengthen the security of the cryptosystem. Finally, we  get the  cipher-image by using the BitXOR operation. The encryption and decryption process is shown in Fig. \ref{fig:encryp-decryp}.  In the following subsections \ref{sec-sub-key-encryp}-\ref{sec-sub-encryption-algorithm} we   demonstrate the encryption/decryption scheme in more detail. 
\subsection{Generation of key  for image encryption}\label{sec-sub-key-encryp}
We generate the key vectors by the chaotic solutions of the system of Eqs. \eqref{E-m}-\eqref{n-m}.  When the chaos is established in the master laser, we  take all the values of $E_{\pm}$ and $N$   sequentially  as vectors to be used in the encryption. Some of these data may be of complex numbers which render   the encryption algorithm more difficult to construct. However, we  represent   the field vectors $E_{\pm}$ and the density vector $N$ in terms of  the chaos sequences are as follows:
\begin{itemize}
\item $E^{+}_{vec}$ = $E^{+}_{1}$,$E^{+}_{2}$,$E^{+}_{3}$,...,$E^{+}_{j}$.
\item $E^{-}_{vec}$ = $E^{-}_{1}$,$E^{-}_{2}$,$E^{-}_{3}$,...,$E^{-}_{j}$.
\item $N_{vec}$ = $N_{1}$,$N_{2}$,$N_{3}$,...,$N_{j}$, where $j$ is the number of the iteration. 
\item  $K^{1}_{i} \longleftarrow$ (round(abs($E^{+}_{i})\times10^{10}$)) mod(256).
\item  $K^{2}_{i} \longleftarrow$ (round(abs$(N)\times10^{10}$)) mod(256).
\item Generate the key by means of the two sequences of integers ${K^1_i}$ and ${K^2_i}$   as   $K_{yi} \longleftarrow$  (abs($K^{1}_{i}$)-floor( abs($K^{2}_{i}$))$\times10^{10}$, mod $256$), 
where  the numbers of elements of $K_{yi}$ vectors are equal to the  lengths of the vectors $E_{+}$ and $N$. 
\end{itemize}
It is to be mentioned that    in our encryption scheme, we take the image   as vectors which are much lower in number than the key vectors.    So,  we   take the key vectors in such a way that all the   vectors form a matrix and each column contains at most $m\times l\times k$ vectors,  i.e.,  $K_{y(m\times l\times k)} \longleftarrow$ reshape $(K_y, m\times l\times k, 1)$. This matrix is now used for encryption of the color image.
\subsection{Algorithm for encryption and decryption} \label{sec-sub-encryption-algorithm}
We present an algorithm for the encryption and decryption of an RGB image using the chaotic data sequence as above. We begin with the shuffling of data using the chaotic time series. We also shuffle the pixel values of the image   so that it becomes totally muddled. This pixel-level permutation  can disrupt the correlation of the adjoining pixels. Furthermore, we  employ a bit-level permutation, to change the bits of each pixel value of the shuffled sequence by a constant matrix.   The processes are stated successively as  follows.
\par 
\textit{\textbf{Shuffling of data using chaotic time series.}} 
 \begin{itemize}
 \item  $A \longleftarrow $ read (Image), where  $A$ is   matrix  (whose elements are  called the voxel values or pixel values) of order $m\times l\times k$. Usually, in an RGB color image, $k=3$  and so $A$ is a three-dimensional matrix.
 \item Convert $A_{m\times l\times k}$ into one array, e.g.,  $T_{m,l,k}=[a_{1,1,1},a_{1,2,1},...,a_{m,l,k}]$.
 \item Having obtained the chaotic sequence  $E^{-}_{vec}$ = $E^{-}_{1}$, $E^{-}_{2}$, $E^{-}_{3}$,...,$E^{-}_{j}$, we   choose $m\times l\times k$ data of $E_{-}$, after discarding first $j_0~(<j)$ number of values as those may not exhibit chaos, by the formula 
$$E'^{-}_{m\times l\times k} = \text{uint} 8(\text{round}(\text{abs}(E^{-}_{i})\times 10^{10}))$$, where $i={j_0, j_0+1,..., j_0+m\times l\times k} $, 
 and arrange  $E'^{-}_{m\times l\times k}$ in ascending order.
\item \textit{Pixel-level permutation}. Operate the BitXor function on each $T_{i,j,k}$ and $E'^{-}_{i}$, so that the pixels get shuffled. Thus, the pixel positions are changed and one obtains the following shuffled sequence  $$Q_{m,l,k}=[{Q_{1,1,1},~Q_{1,2,1},~Q_{1,3,1},...,Q_{m,l,k}}].$$
\end{itemize}
\textit{\textbf{Bit-level permutation.}}
\par
Here, we   change the bits of each pixel value of the shuffled sequence using a constant matrix that is to be formed using the initial conditions and the parameter values which exhibit chaos.
\begin{itemize}
 \item Form the matrix $M$ as: Let  $d_1=\max(m,l,k)$ and $d_2=\min(m,l,k)$. Define
$s_i=\left(x_i+m\times l\times k\right)/\left(2^{16}+m\times l\times k\right)$ for $ i = 1,2,..,r$ with       the  initial condition $x_i$  and   $m_i=\text{uint}8\left(mod\left(s_i\times 10^{16},1\right)\right)$.   Here,  the factor $2^{16}$ is considered to make $s_i$ much smaller than the unity, and  that no recurrence occurs in the  decimal representation of $s_i$. Also, in the definition of $m_i$, $s_i$ is multiplied by $10^{16}$ in order to retain the  values of $m_i$ up to $16$ significant digits (for security reason).      Then
 \begin{equation}
M=
  \begin{bmatrix}
    m_1 & m_2 & m_2 & ... & m_r \\
    ...&...&...&...&...\\
    m_r & m_{1} & m_{2} & ... & m_{r-1}
  \end{bmatrix}.
\end{equation}
\item Divide each sequence of $Q_{m,l,k}$ into a matrix of order $r\times r \times  \left(m\times l\times k\right)/r $,  where $r$ is one of the divisors of $m\times l\times k$.
\item Do bit-level operation between each matrix of order $r\times r$ and the matrix $M$.
\item Repeat the previous step until $\frac{m\times l\times k}{r} $ number of matrices have executed a round of bit-level permutation operation.
\par
Finally,  we get the bit-level  permutation matrix $P_{m\times l\times k} $  of the original image.   
\end{itemize}
\textit{\textbf{Encryption process.}}
\par 
  We use the key vector $K_{y(m\times l\times k)}$ as in  Sec. \ref{sec-sub-key-encryp}  to encrypt the permutation vector $P_{m\times l\times k}$ using the formulas, given by,\\
   $D_{1}=\text{mod}((P_1 \oplus D_0), 256) \oplus \text{mod}((Q_1 \oplus K_{y1}),256)$, \\ 
  $D_{i}=\text{mod}((P_i \oplus D_{i-1}), 256) \oplus \text{mod}((Q_i \oplus K_{yi}),256)$,\\ 
where $D_0$ is a constant vector of order ${m\times l\times k}$,  $i=2,...,m\times l\times k$. We then repeat the previous process(es) until  the full encryption is done, and we get the cipher image as  $ CI\longleftarrow$ reshape$(D,m,l,k)$ 
\par
\begin{figure*}[hbtp]
 \centering
 \includegraphics[scale=.4]{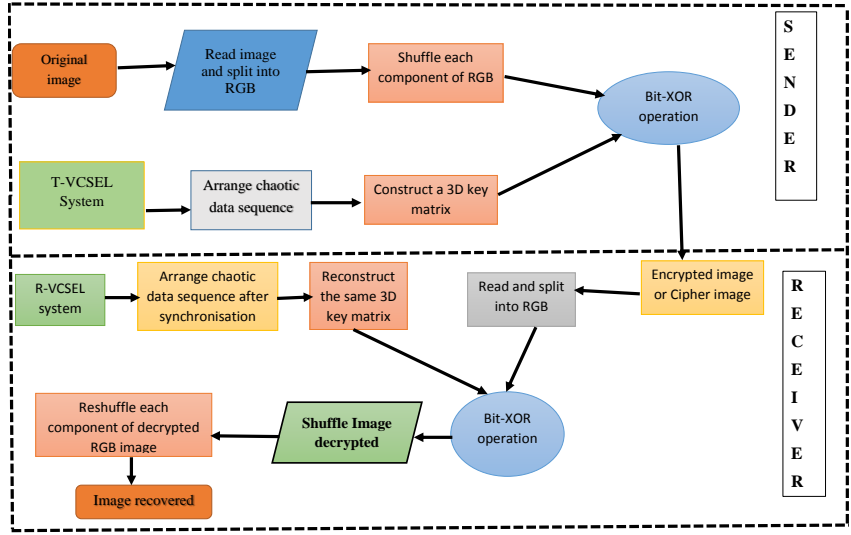}
 \caption{A pictorial diagram showing the encryption and decryption of images  through chaotic medium. }
 \label{pictorial-diagram}
 \end{figure*}

 
Thus, following  the above processes one can encrypt an image using the chaos vectors $E_{\pm}, ~N$ and $n$. In Fig. \ref{pictorial-diagram} we show  how a VCSEL is used for encryption and decryption of images or data through synchronization of master and slave lasers.  
%%%%%%%%%%%%%%%%%%%%%%%%%%%%%%%%% Recovering data through synchronization   %%%%%%%%%%%%%%%%%%%
 %%%%%%%%%%%%%%%%%%%%%%%%%%%%%%%%%%%%%%%%%%%%%%%%%%%
Next, we now demonstrate  the decryption algorithm   as  follows:
\par
\textit{\textbf{Decryption process.}}
\par
  The decryption process is the reverse process of  encryption. Firstly,  as soon as the synchronization of the two coupled VCSELs is achieved, the receiver  obtains the cipher image, as well as,   the initial conditions and the parameter values (those in the process of encryption) to generate  the key. Here, the   the key for decryption  can be generated from the slave laser by  the same way as in the  process of  encryption.  Secondly, having obtained the cypher image and the key vectors  we should repeat the same process as in  the processes of shuffling the data and bit-level permutation.
 \par
  Finally, we should use the BitXOR operation between the cypher image  and the generated key vectors to recover the image data vectors, i.e., the decrypted cipher image (DCI) using the formulas as\\
  $D'_{1}=mod((P'_1 \oplus D'_{0}),256) \oplus mod((Q'_1 \oplus K_{y1}),256)$,\\
 $D'_{i}=mod((P'_i \oplus D'_{i-1}),256) \oplus mod((Q'_i \oplus K{yi}),256)$,\\ where $i=2,...,m\times l\times k$; $D'_i$ stands for the inversion of $D_i$, $P'_i$ is the bit-level permutation vector which is generated   from $CI$ and $Q'_i$ is the pixel shuffling vector of $CI$. We repeat the process(es) until the decryption  of the cipher image is completed, and we recover  the original image.
 After  the original image $A$ is recovered, reshape it as  $A \longleftarrow $ reshape $(D'_{m,l,k})$.
 %%%%%%%%%%%%%%%%%%
\section{Security analysis of  encrypted image}\label{sec-security-analysis}
In this section we will test and analyze  our proposed cryptosystem whether it is resistant to statistical and differential attacks. In fact, the cipher image (CI)   remains secured under the  process of data transmission because the secrete key, which is based on the initial conditions and the range of parameters,   securely passes through the communication channel during the synchronization of the couple system of VCSELs. The security analysis is given in  subsections \ref{sec-sub-sensivity-analys}-\ref{sec-sub-qqplot} which include the   analysis of key space,  differential attack resistance, histogram, correlation coefficient, quantile-quantile plot and the entropy analysis. To this end, we consider a $400\times300$ image ``Autumn Leaves" as in Fig. \ref{fig:encryp-decryp}(a).  It is  shown that even a small    change of the key development can't help recover images though the synchronization, i.e.,    the encryption is free from any brute force attack.
\subsection{Key space and sensitivity analysis}\label{sec-sub-sensivity-analys}
The numerical   set of values of $E_{\pm}$ and $N$, which exhibit chaos  under suitable   initial conditions and parameters  (as in  Sec. \ref{sec-model}) are arranged to establish the key vectors.  We carry out the analysis of the key space attack resistance and the sensitivity analysis to get more secured  data against any brute force attack effectiveness. The key space mainly consists of a wide number of key vectors which are generated by   the solutions of the  dynamical system of Eqs. \eqref{E-m}-\eqref{n-m} for the master laser which exhibit  chaos.     Each key vector is so generated   that  its elements   lie between $0-225$ or in between $0-511$ as per   the size of the image (e.g., $256\times 256\times 3$, $400\times300$, $512\times 512\times 3$ etc.). These key vectors are then used for encoding the  RGB image by the proposed algorithm as in Sec. \ref{sec-encryption}.   Here, any wrong key representation for decrypting the image gives an incorrect set of pixel values which is again a blur image. For key space analysis, we   consider the initial condition that is used to generate chaos in the VCSELs and  the key matrix to be of the order four for which we have $(10^{16})^4$ possible key elements. Again in Sec. \ref{sec-sub-key-encryp} for the generation of key, we have $(10^{10}\times 10^{10})=10^{20}$ number of possible data set which represents the key elements. Also, in the process of encryption and key representation using the BitXor operation ($8$ bit operation), each layer of matrices has order $4$. So,  if $N_0$ denotes the total number of iterations in the encryption and decryption processes, then the total no of possible key  is $N_0\times 2^{8}\times 10^{64}\times 10^{20}\approx N_0\times 2^{288}$, which is large enough to determine the key i.e. it is free from brute force attack. 
\par
The dynamical systems of VCSELs are sensitive to initial conditions, a small change of which can result  into a different kind of solution, and hence an incorrect image or blur image even if the synchronization is achieved. The initial conditions  by which the system exhibits chaos   are mainly used  for encoding and decoding the dataset. However, if  a small change of values of  $E_{\pm}$ occur, e.g., from $E_{\pm}=0.001$ to $E_{\pm}=0.0015$, a different   kind of key will be generated by which   the particular image or information may not be recovered. 
\par
The steps for the generation of key as in Sec. \ref{sec-sub-key-encryp} should also be followed in order. Otherwise, any change in between the steps results into an incorrect key, i.e., an incorrect representation of the image. In this case,  the position of each pixel values of the image  matrix will be changed and the effectiveness of  displaying the image will result into a blur image. Thus,  it may not be easy to recover the image even if  one   guesses a value of the key, i.e., the image will be transmitted to the receiver section secretly. 
%%%%%%%%%%%%%%% Pictorial diagram%%%%%%%%%%%%%%%%%
 \begin{figure*}[hbtp]
 \centering
 \includegraphics[scale=.6]{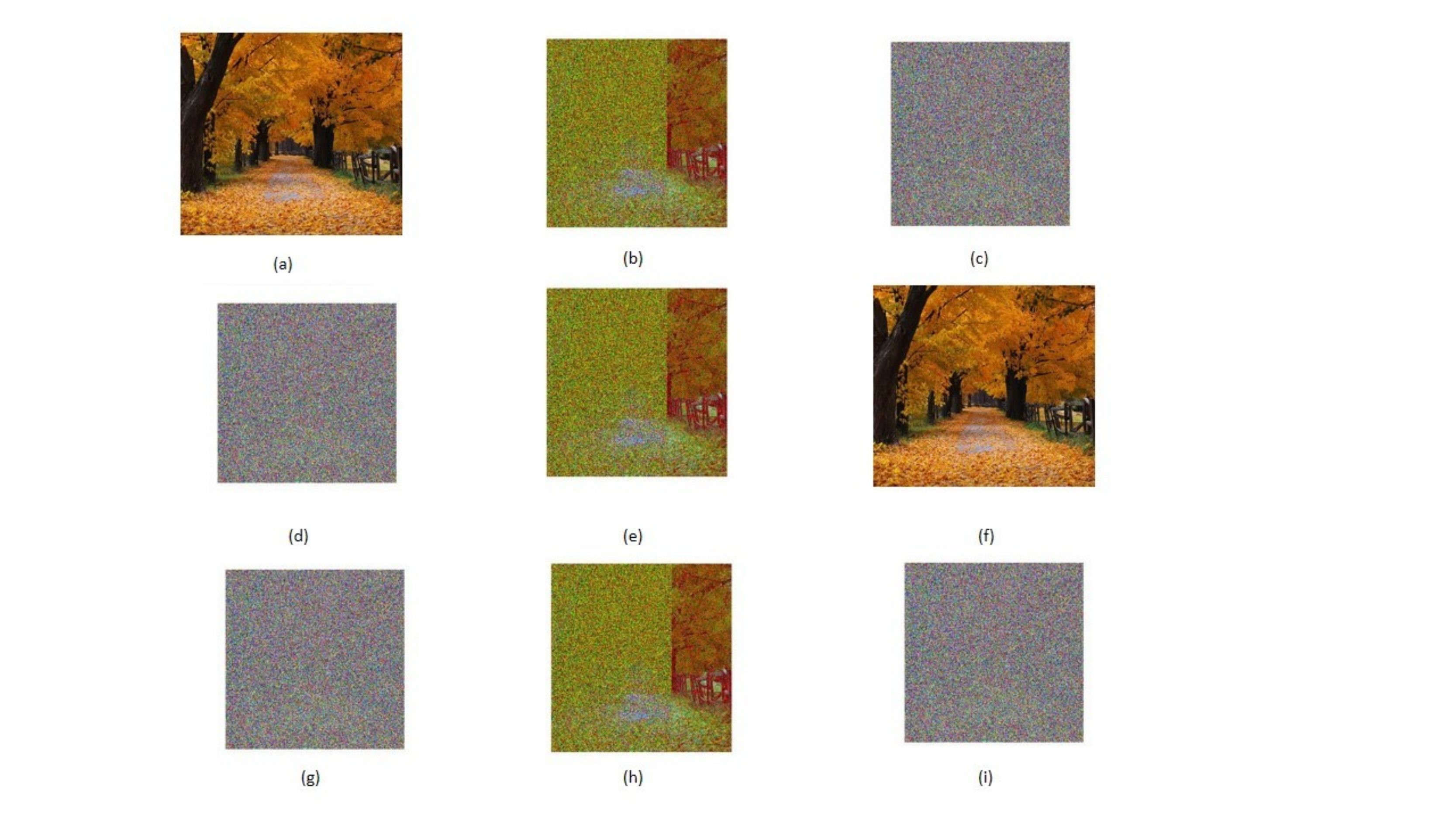}
 \caption{ Encryption and decryption using the correct and wrong keys. Subfigures (a)-(c) indicate the encryption process.  While subfigures (d)-(f) show the decryption process using the right key, subfigures  (g)-(i) correspond to the incorrect image or blur image  using a wrong key.}
 \label{fig:encryp-decryp}
 \end{figure*}
 %%%%%%%%%%%%%%%%%%%%%%%%%%%%%%%
  In the receiver section, the key vectors are generated for decryption of the image through the synchronization of coupled VCSELs. Here, one must note that the generation of the key matrix must be only once throughout the   synchronization.   Fig. \ref{fig:encryp-decryp}  shows the decryption of the image ``Autumn Leaves" with correct and incorrect keys by the process as described above.  
  \subsection{Differential attack resistance}\label{sec-sub-attack-resis}
 To be resistant to a differential attack, a good cryptosystem must ensure that any small change or modification in the plain image  results into a significant difference in the cipher image.   In the formation of cipher image, since the pixel values of the plane image are taken in a proper order it is very difficult for hackers to recover the image even if  they can make a small change of the pixel values for which  a different cipher image is created thereby resisting the differential attack efficiently. To test the effects of only one change of pixel value from the plain image to the cipher image, we   introduce two common measures, one the number of pixels change rate (NPCR) and other the unified average changing intensity (UACI). They are defined as follows:
 \begin{equation}
  NPCR =\dfrac{ \sum_{i,j}^{m,n}D(i,j)}{m\times n} \times 100, 
\end{equation}  
   where $D(i,j)$ represents the change of the pixel values from the plain image to the  cipher image due to the encryption process,i.e., 
   \begin{equation}
   D(i,j)= \left\lbrace \begin{array}{ccc}  
      0 & when &  P(i,j)=CI(i,j) \\
     1  & when &  P(i,j)\neq CI(i,j). 
   \end{array} \right.
    \end{equation}
As an illustration, we show an example of how the pixel values are changed from the plane image to the cypher image in the process of encryption.
\begin{equation}
\begin{array}{cc}
  P(1,15,1)= 210, & CI(1,15,1)=17 \\
   P(4,4,1)=185,& CI(4,4,1)=3\\
  P(200,5,2)=25, & CI(200,5,2)=2\\
   P(100,100,2)=81, & CI(100,100,2)=243\\ 
   P(10,18,3)=31, & CI(10,18,3)=224\\
   P(25,100,3)=37, & CI(25,100,3)=238,
 \end{array}
 \end{equation}
 where the matrix $P~(CI)$   corresponds to the  plain (cypher) image and their values are in the range $0-255$, i.e., integer modulo $256$. As the  pixel values are changed, we use UACI to determine the average intensity of the difference between the original and the encrypted image, where
 \begin{equation}
  UACI= \dfrac{1}{m\times n} \sum_{i,j}^{m,n} \dfrac{|P(i,j)-CI(i,j)|}{255} \times 100. 
 \end{equation}
Next, we show the results of NPCR and UACI in  Table \ref{table-1}.
\begin{table}[]
 \begin{center}
\begin{tabular}{ |c|c|c|c| } 
 \hline
   & R & G & B \\ 
   \hline
  NPCR & 99.73 & 99.78 & 99.74 \\ 
  \hline
 UACI & 33.23  & 33.27 & 33.14 \\ 
 \hline
  \end{tabular}
\end{center}
\caption{The measures of the number of pixels change rate (NPCR) and   the unified average changing intensity (UACI) for an RGB image are shown.} 
\label{table-1}
\end{table} 
 It is evident that the scheme has high NCPR values with satisfactory UCAI values giving a resistant to    any differential attack.
 \subsection{Comparison of times for image encryption with some existing schemes} \label{sec-sub-time-comparison }
 The proposed algorithm is tested using Matlab $2016$a in a laptop with $2.20$ GHz processor Intel Core $i3-2328$ CPU, $4$GB RAM, and Windows $8$, $64$-bit operating system. To this end, we consider the RGB Lena.jpg image of different sizes ($256\times 256\times 3$, $512\times 512\times 3$ and $1024\times 1024\times 3$). The average encryption/decryption times (in seconds) using different schemes proposed in different articles are compared with our   algorithm as given in   Tables \ref{table-time1} and \ref{table-time2}. It is seen that our scheme  provides   less time to encrypt/decrypt an image than the existing ones.  
\begin{table}[!h]
\begin{center}
\begin{tabular}{|c|c|c|c|c|c|}
\hline
  RGB  & Time  & Time &   Time &   Time \\
  (Image & (Our  & (Ref. & (Ref. & (Ref. \\
   size)& scheme)& \cite{assad2016})&  \cite{socek2005}) &  \cite{subramanyan2011}) \\
 %Image size& (Our scheme)& (Ref. \cite{assad2016})& (Ref. \cite{socek2005})& (Ref. \cite{subramanyan2011})\\
\hline
$256\times 256\times 3$   & 1.01 & 1.05 & 1.56 & 2.85 \\
\hline
$512\times 512\times 3 $   & 1.83 & 2.01 & 2.34 & 4.43\\                             
\hline 
$1024\times 1024\times 3$  & 4.6 & 5.06 & 5.56 & 7.81 \\                   
\hline
 \end{tabular}
\end{center}
\caption{Different encryption times (in seconds) of an RGB (\textit{Lena.jpg}) image using different schemes are shown to ensure that our proposed scheme provides relatively less time to encrypt and decrypt an  image. } 
\label{table-time1}
\end{table}
%%%%%%%%%%%%%%%%%%%%%%%%%

\begin{table}[!h]
\begin{center}
\begin{tabular}{|c|c|c|c|c|c|}
\hline
 RGB  & Time  & Time &   Time &   Time \\
  (Image & (Our  & (Ref. & (Ref. & (Ref. \\
   size)& scheme)& \cite{assad2016})&  \cite{socek2005}) &  \cite{subramanyan2011}) \\
 %Image size& (Our scheme)& (\cite{assad2016})& (\cite{socek2005})& (\cite{subramanyan2011})\\
\hline
$256\times 256\times 3$   & 1.0625 & 1.0841 & 2.36 & 3.25 \\
\hline
$512\times 512\times 3 $   & 4.17 &  4.393 & 6.24 & 8.43\\                             
\hline 
$1024\times 1024\times 3$  & 17.13 & 17.356 & 35.246 & 48.321 \\                   
\hline
\end{tabular}
\end{center}
\caption{Different encryption times (in seconds) of an RGB (\textit{Autumn leaves.jpg}) image using different schemes are shown to ensure that our proposed scheme provides relatively less time to encrypt and decrypt an  image. }  
\label{table-time2}
\end{table}  
%%%%%%%%%%%%%%%%

\subsection{Histogram analysis} \label{sec-sub-histogram}
\begin{figure*}[hbtp]
 \centering
 \includegraphics[scale=.3]{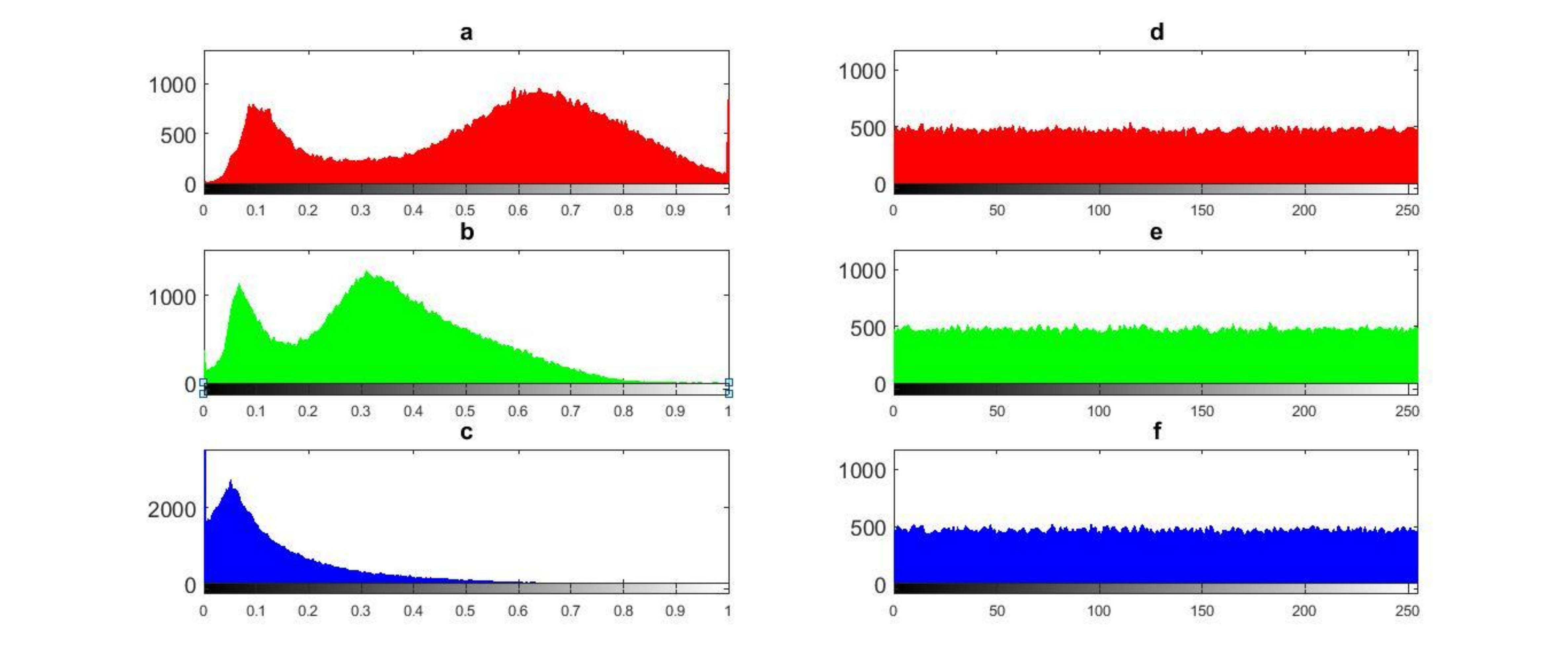}
 \caption{Histogram analysis: Subfigures (a), (b) and (c), respectively,  correspond to the R, G and B components of the original image, and   (d), (e) and (f) are those   for the cipher image.}
 \label{fig:histogram}
 \end{figure*}

A histogram analysis corresponding to an image is mainly concerned with the distribution of the pixel
(intensity) values within the image.  Any encryption scheme is said to be secured if the encrypted image can have a uniform histogram to resist any statistical attacks.  The histograms of the original RGB image ``Autumn Leaves" and the corresponding cipher image  are shown in Fig.  \ref{fig:histogram}.  
\par We note that for an RGB image,   the ordered pixels are scrambled or manipulated and their  distribution is studied with the help of a chart   representing the distribution of the pixels in the range $0-255$.  The $q$-th gray level $l_q$ of a gray image  is represented by a function  $\text{hist}(l_q)=n_q$, where   $n_q$ denotes the number of pixels in the image.  In the diffusion phase, the positions of the image elements are then shuffled so that the statistical information of the original image remains unaltered. However, an
additional layer of security will disguise the desired information, as reflected in the histogram  (Fig. \ref{fig:histogram}).  It is also clear from Fig. \ref{fig:histogram} that the gray-scale values of the encrypted image (see the right panels d to f) are uniformly distributed over the interval $[0~255]$, which is  significantly different from that of the original image (see the left panels a to c).  Thus, an attacker will be unable to infer any statistical information required to decode from the scheme. From the results of the
pixel intensity distributions, it can be ascertained that the scheme also possesses good confusion properties.   
%%%%%%%%%%%%%%%%%%%%%%%%% Histogram%%%%%%%%%%%%%%%%%%%%%%%%%%
 %%%%%%%%%%%%%%%%%%%%%%%%%%%%%%%%%%%%%%%%%%%%%%%%
 \subsection{Correlation analysis} \label{sec-sub-correl}
\begin{figure*}[hbtp]
 \centering
 \includegraphics[scale=.4]{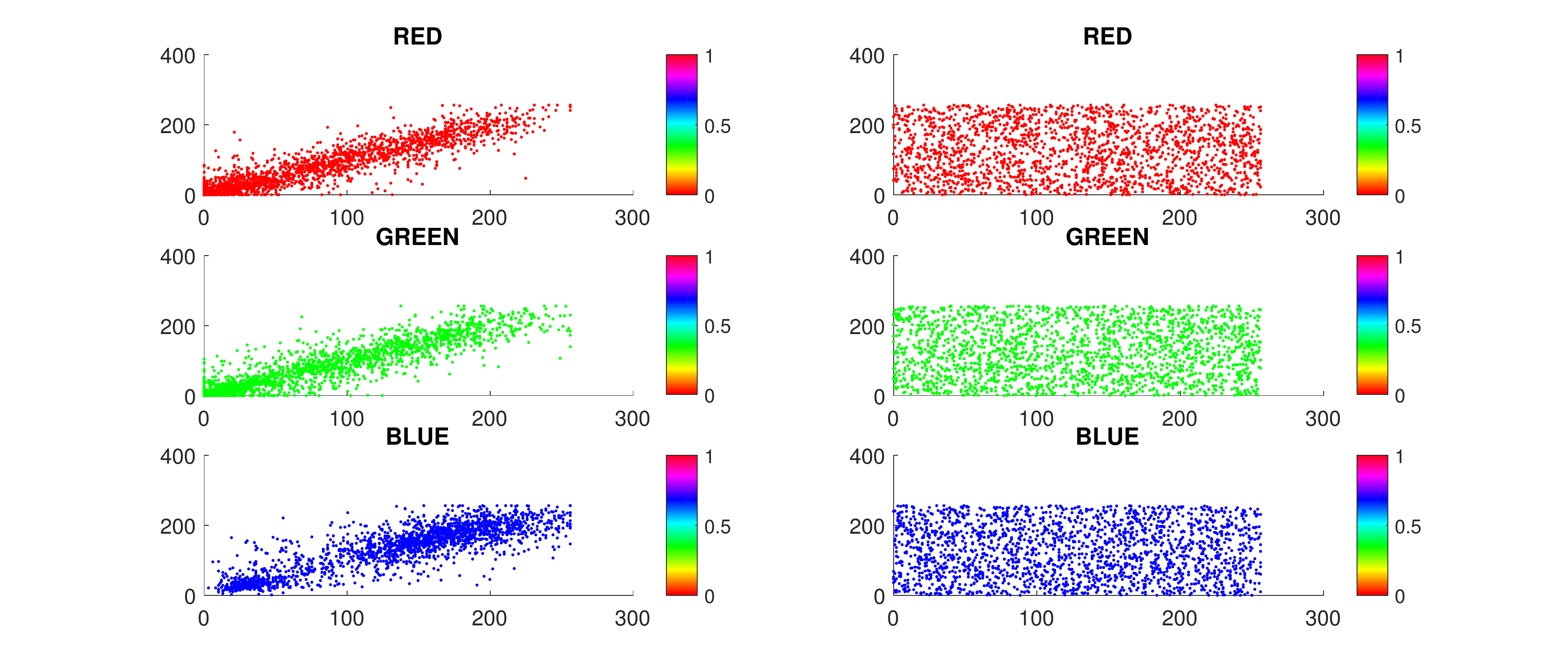}
 \caption{Scatter plots of the correlation coefficients of the adjacent pixels of   the original RGB image (left panel) and  the  cipher image (right panel). The subplots a, b and c (d, e and f) are corresponding to the R, G, and B components of the original image (cipher image).}
 \label{fig:scatter-corr}
 \end{figure*}

The adjacent pixels of the original image are highly correlated while distributed along the horizontal (H), vertical (V) and diagonal (D) directions. However, an ideal encryption algorithm ensures that for  the encrypted image  the correlation coefficients of the adjacent pixels is nearly zero  to resist any statistical attack.  The correlations of the adjacent pixels in the plain and cipher images are  analyzed and compared as shown in Fig. \ref{fig:scatter-corr}. 
\par
To define the correlation coefficients, we first define  the covariance between a pair of  pixel values $x$ and $y$ as
$Cov(x,y) = E[(x-E(x))(y-E(y))]$. Then  the corresponding correlation coefficient is given by
\begin{equation}
\rho_(xy)=\dfrac{Cov(x,y)}{\sigma(x)\sigma(y)},~ ~\sigma(x),~\sigma(y)\neq0,
\end{equation}
where $E(x)$ and $E(y)$ are  the means, and $\sigma(x)$ and $\sigma(y)$ are the standard  deviations of the distribution of the pixel values  which range from  $0$ to $255$. The adjacent pixel values  are placed   horizontally, diagonally and vertically.   The values of $\rho$ are computed for  the RGB  image ``Autumn Leaves"and the cipher image as given in   Tables  \ref{table-rgb} and \ref{table-cipher}. 
%%%%%%%%%%%%%%%%%%%%%%%%%%%%%%%%%%%%%%%%%%%%%%
\begin{table}[!h]
\begin{center}
\begin{tabular}{|c|c|c|c|}
\hline
$\rho$ & R & G & B \\
\hline
H & 0.9041 & 0.8109 & 0.7347 \\
\hline
V & 0.8875 & 0.7748 & 0.6725 \\
\hline 
D & 0.8551 & 0.7194 & 0.6094\\
\hline
\end{tabular}
\end{center}
\caption{Correlation coefficients of  a pair of adjacent pixel values  of the plain image ``Autumn Leaves" while distributed along the horizontal (H), vertical (V) and diagonal (D)     directions. }
\label{table-rgb}
\end{table}
%%%%
\begin{table}[!h]
\begin{center}
\begin{tabular}{|c|c|c|c|}
\hline
$\rho$ & R & G & B  \\
\hline
H & -0.0008 & -0.0013 & 0.0056\\
\hline
V & 0.0006 & 0.0017 & -0.0032\\
\hline
D & 0.0036 & 0.0007 & -0.0013 \\
\hline
\end{tabular}
\end{center}
 \caption{Correlation coefficients of  a pair of adjacent pixel values,  corresponding to the cipher image of ``Autumn Leaves", while distributed along the horizontal (H), vertical (V) and diagonal (D) directions.}
 \label{table-cipher}
 \end{table}  
 %%%%%%%%%%%%%%%%%%%%%%%%%%%%%%%%%%%%%
From the computed values it is observed that the correlation coefficients for the cipher image are very low 
i.e., close to zero. This is also evident from the  scatter diagrams for the RGB   and the cipher images as in Fig. \ref{fig:scatter-corr}.  We find that   it is hard to correlate between the plain   and the cipher images. For an encryption scheme to be efficient, it is imperative that the correlation coefficient between the adjacent pixels be minimal for the cipher image. This is what we have also obtained in the present analysis.     
 %%%%%%%%%%%%%%%%%%%%%%%%%%%%%%%%%%%%%%%%%%%%%%%%%%%%%%%%
 %%%%%%%%%%%%%%%%%%%%%%%%%%%%%%%%%%%%%%%%%%%%%%%%%%%%
 \subsection{Q-Q plot and comparison } \label{sec-sub-qqplot} 
\begin{figure*}[]
    \centering
            \includegraphics[width=6.5in,height=3in]{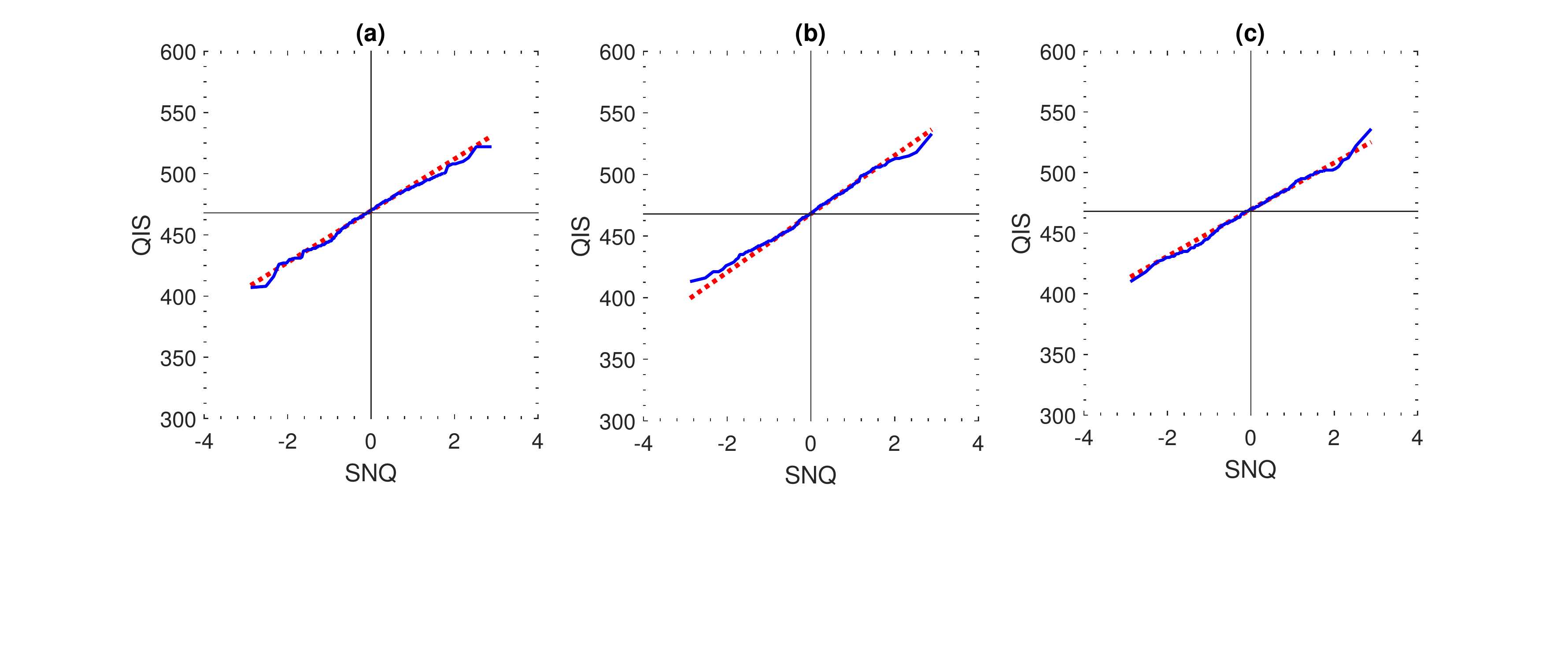} 
    \caption{Q-Q plots  of the (a) Red    (b) Green and (c) Blue components  of the encrypted RGB image using the   algorithm as in Sec. \ref{sec-encryption}. The acronyms SNQ and QIS, respectively, stand for the standard normal quantile  (red dotted lines) and the quantile  of input sample  (blue dashed lines).  }
    \label{fig:qq-plot}
\end{figure*}

For a quantitative analysis of the red, green and blue components of the encrypted RGB   image ``Autumn Leaves" as in Fig. \ref{fig:encryp-decryp}(a), we plot the quantiles of input sample (QIP) against the standard normal quantile (SNQ), i.e.,   the quantile-quantile (Q-Q) plot, which is  used to check whether the points of the sample data set fall  approximately along a reference line (e.g., the $45$-degree reference line).    In general, the basic idea is to compute  theoretically the expected value for each data point based on the distribution under consideration. If the data, indeed, follows the assumed distribution, then the points on the Q-Q plot will fall approximately on a straight line.  From Fig. \ref{fig:qq-plot}, it is evident that   the data points of the  red, green and blue fall approximately along the reference line.  This approximate linearity of the points suggests that the data are normally distributed. 
%%%%%%%%%%%%%%%%%%%%%%%%%%%%%%%%%%%%%%%%%%%%%%%%%%%%%%%%%%%%%%%%%%%%
 \begin{figure*}[]
    \centering
            \includegraphics[width=6.5in,height=3in]{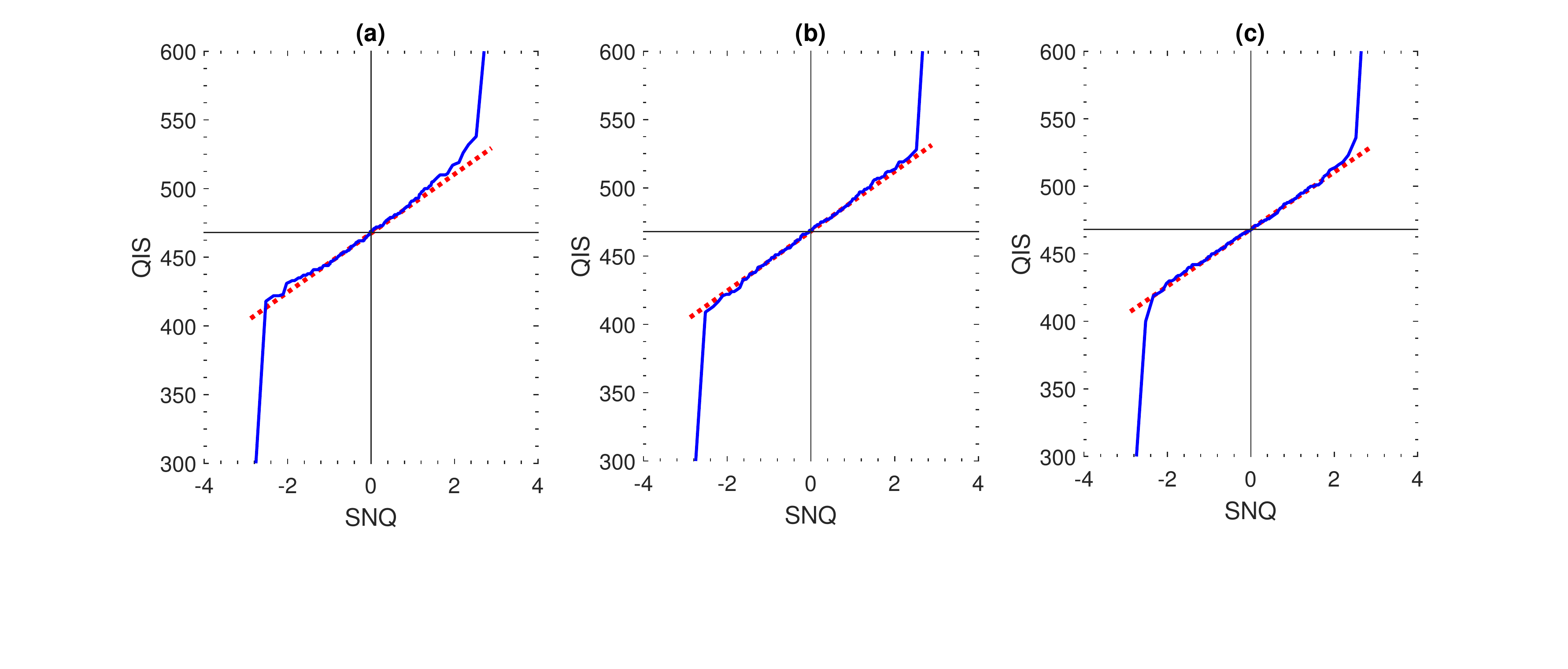}       
    \caption{Q-Q plots of the (a) Red    (b) Green and (c) Blue components  of the encrypted RGB image using the algorithm as in Ref. \citep{li2017}. The acronyms SNQ and QIS, respectively, stand for the standard normal quantile (red dotted lines) and the quantile  of input sample (blue dashed lines).  }
    \label{fig:qq-plot-comp}
\end{figure*}
%%%%%%%%%%%%%%%%%%%%%%%%%%%%%%%%%%%%%%%%%%%%%%%%%%%%%%%%%%%%%%%%%%%
\par We ought to mention that our present encryption algorithm significantly modifies that proposed by   Li \textit{et al.} \cite{li2017}.  The modification is mainly due to the use of the BitXor function in the bit-level permutation.    The necessity for such a modification  is based on some  reasons, namely,
 \begin{itemize}
 \item Firstly, using the algorithm   of   Li \textit{et al.} \cite{li2017},  the matrix multiplication between the $4\times 4$ constant matrix and the pixel-shuffled matrix  of order $4\times 4 \times \frac{MN}{4}$  in the  bit-level permutation  takes  much longer time than using our proposed  algorithm.  Furthermore,   the  encryption/decryption algorithm of   Li \textit{et al.} \cite{li2017} may not be a good resistant to any time attack or any kind of brute force attack.    
 \item Secondly,   Li \textit{et al.} \cite{li2017} considered a  two-dimensional digital image matrix, however, in the present theory, we have considered an  RGB image which has more pixel values than the  digital image matrix.  So, in the diffusion process, a simple scalar addition of the bit-level permutation matrix with the chaotic data set   makes  the decryption process much more complicated, however, the bit-wise XOR operation, as in the present algorithm, causes no such problem in addition. Thus, the encryption scheme in Ref. \citep{li2017} may be safe, but may not be much secured compared to our present algorithm.
 
\end{itemize} 
In order to compare our proposed algorithm  with that in Ref. \citep{li2017},  and to show why our algorithm provides better security for encryption and decryption, we consider the RGB image ``Autumn Leaves" as in Fig. \ref{fig:encryp-decryp}(a). The results for the Q-Q plots are displayed in Fig. \ref{fig:qq-plot-comp}. Comparing the results in Figs. \ref{fig:qq-plot} with  \ref{fig:qq-plot-comp}, one can conclude that the encryption algorithm in Ref. \citep{li2017}  has some limitations as evident from the singularities at some points of the data set, which are, however, removed by the modified algorithm as proposed in the present work.    
   
%%%%%%%%%%%%%%%%%%%%%%%%%%%%%%%%%
\subsection{Entropy Analysis}
\begin{figure*}[]
    \centering
    \begin{subfigure}[]%{0.5\textwidth}
        \centering
        \includegraphics[height=1.2in]{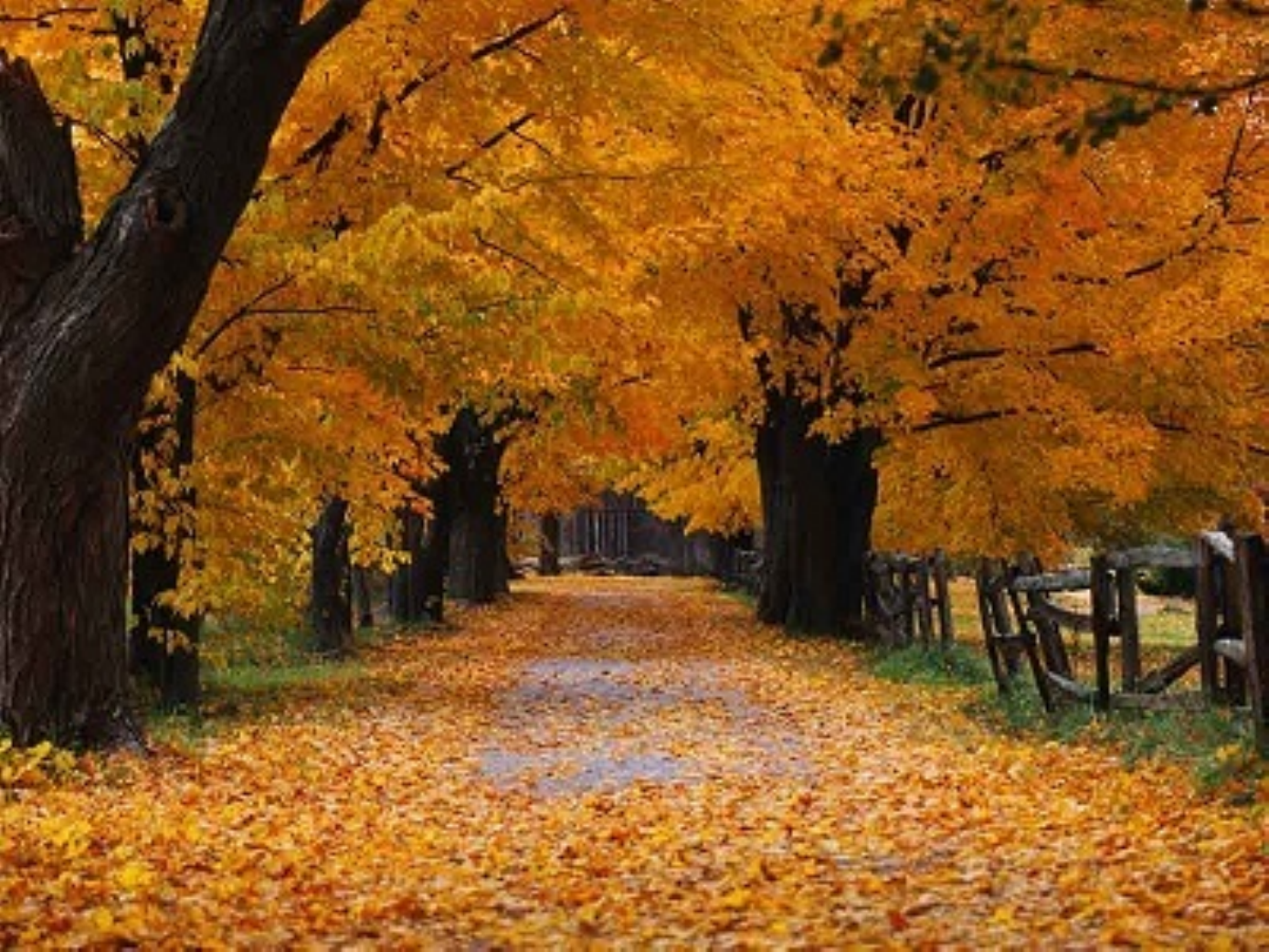}
       %\caption{fig 1}
       \label{fig:corr_roop}
    \end{subfigure}%
    ~ 
    \begin{subfigure}[]%{0.5\textwidth}
        \centering
        \includegraphics[height=1.2in]{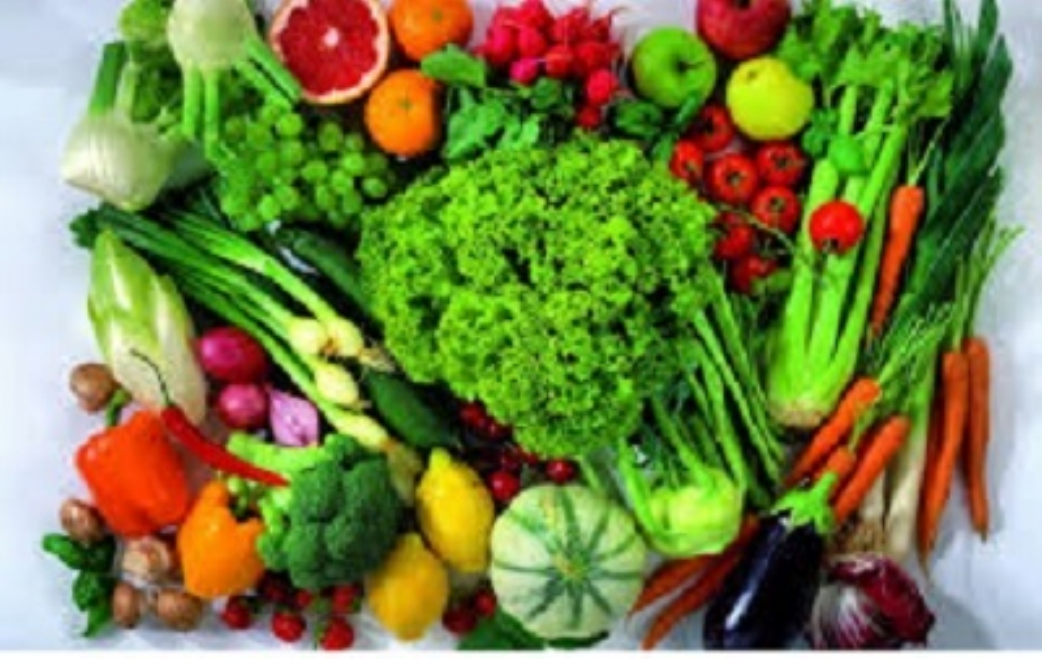}
        %\caption{encrypted green component qqplot}
        \label{fig:corr_images}
    \end{subfigure}
        \begin{subfigure}[]%{0.5\textwidth}
        \centering
        \includegraphics[height=1.2in]{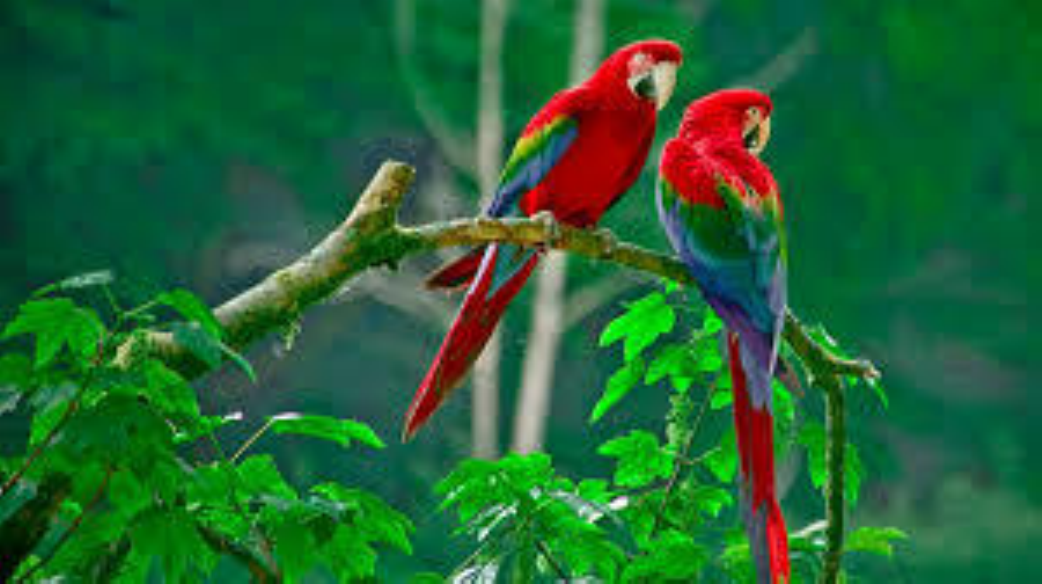}
        \label{fig:corr_download}
    \end{subfigure}
    \caption{Three different images for which the entropy values are calculated. The corresponding entropy values are given in Tables  \ref{table-original-image} and \ref{table-cypher-image}.}
    \label{fig:entropy-encrypt-image}
\end{figure*}

Here, we present an another statistical measure of uncertainty for the RGB image. As per the Shannon  entropy \cite{shannon1948}, it is the expected value of the information contained in an image, and is defined by   
\begin{equation}
H(X)=E[I(X)]=-E[\ln(P(X))], \label{eq-entropy1}
\end{equation}
where $E$ is the expected value operator and $I$ is the information content of the random variable $X$. Equation \eqref{eq-entropy1} can also be written as
\begin{equation}
H(x)=\sum_{i=0}^{255}P(x_i)I(x_i)=-\sum_{i=0}^{255}P(x_i)\log_2P(x_i). \label{eq-entropy2}
\end{equation}
We assume that there are $256$ values of the information source in Red, Green and Blue colors of the image with the same probability. We can get the perfect entropy $H(X) = 8$, corresponding to a truly random sample  \cite{stoyanov2015}. The information entropy of Red, Green and Blue   colors of the plain and their corresponding encrypted images are computed and displayed in Tables \ref{table-original-image} and \ref{table-cypher-image}. To this end, we choose three   different images  for the three different samples  as in  Fig. \ref{fig:entropy-encrypt-image} (one of which is the ``Autumn Leaves" image). From Table  \ref{table-cypher-image}, it is evident that the entropy values corresponding to the encrypted RGB image are close to $8$ as expected  for a true random sample. 
%%%%%%%%%%%%%%%%%%%%%%%%%%%%%%%%%%
 
%%%%
\begin{table}[!h]
\begin{center}
\begin{tabular}{|c|c|c|c|}
\hline
Image index & R & G & B  \\
\hline
Fig. \ref{fig:entropy-encrypt-image}(a) & 7.7644 & 7.4035 & 6.0690\\
\hline
Fig. \ref{fig:entropy-encrypt-image}(b) & 7.9450 & 7.9556 & 7.0454\\
\hline
Fig. \ref{fig:entropy-encrypt-image}(c) & 5.6564 & 6.8519 & 6.6589 \\
\hline
\end{tabular}
\end{center}
 \caption{Entropy values corresponding to the original images as in Figs. \ref{fig:entropy-encrypt-image}(a)-\ref{fig:entropy-encrypt-image}(c).}
 \label{table-original-image}
 \end{table}
 %%%%
 \begin{table}[!h]
\begin{center}
\begin{tabular}{|c|c|c|c|}
\hline
Image index & R & G & B  \\
\hline
Fig. \ref{fig:entropy-encrypt-image}(a) & 7.9908 & 7.9906 & 7.9907\\
\hline
Fig. \ref{fig:entropy-encrypt-image}(b) & 7.9903 & 7.9908 & 7.9902\\
\hline
Fig. \ref{fig:entropy-encrypt-image}(c) & 7.9878 & 7.9874 & 7.9891 \\
\hline
\end{tabular}
\end{center}
 \caption{Entropy values corresponding to the cipher images of Figs. \ref{fig:entropy-encrypt-image}(a)-\ref{fig:entropy-encrypt-image}(c).}
 \label{table-cypher-image}
 \end{table}
 %%%
 %%%%%%%%%%%%%%%%%%%%%%%%%%%%%%%%%%%%%%%%%%%%%%%%%%%
\section{Conclusion}
We have investigated the chaos and  synchronization properties of electric field polarizations and the carrier population densities in    a free-running vertical-cavity surface-emitting laser (VCSEL). A two-way coupling of   master and slave lasers in  VCSELs is considered, which is shown to exhibit hyperchaos and  synchronization with a high level of similarity.    The coupled VCSELs are then used as a sender and  a receiver for the communication of image or data.  We have also  proposed a modified chaos-based image encryption algorithm using the pixel- and bit-level permutations which is robust, faster and simpler in encryption/decryption compared to many other chaos-based cryptosystems \cite{assad2016,socek2005,subramanyan2011}. The  new cryptosystem  is analyzed and compared with a recently proposed one \cite{li2017}.    It is shown (by  the  security analysis)  that  the proposed cryptosystem is resistant to various types of attacks and  is efficient for secure communications in nonlinear optical media. 
 %%%%%%%%%%%%%%%%%%%%%%%%%%%%%%%%%%%%%%%%%%%%%%%%%%%%%%%%%%%%%%%
\section*{Acknowledgements} This work  was supported by UGC-SAP (DRS, Phase III) with  Sanction  order No.  F.510/3/DRS-III/2015(SAPI),   and UGC-MRP with F. No. 43-539/2014 (SR) and FD Diary No. 3668.
%%%%%%%%%%%%%%%%%%%%%%%%%%%%%%

\end{document}